# A Fuzzy Clustering Based Approach for Mining Usage Profiles from Web Log Data

Zahid Ansari[1], Mohammad Fazle Azeem[2], A. Vinaya Babu[3] and Waseem Ahmed[4]

[1,4]Dept. of Computer Science Engineering
[2]Dept. of Electronics and Communication Engineering
P.A. College of Engineering
Mangalore, India
[1]zahid.ansari@acm.org
[2]mf.azeem@gmail.com
[4]waseem@computer.org

[3]Dept. of Computer Science Engineering
Jawaharlal Nehru Technological University
Hyderabad, India
dravinayababu@jntuh.ac.in

*Abstract*— The World Wide Web continues to grow at an amazing rate in both the size and complexity of Web sites and is well on it's way to being the main reservoir of information and data. Due to this increase in growth and complexity of WWW, web site publishers are facing increasing difficulty in attracting and retaining users. To design popular and attractive websites publishers must understand their users' needs. Therefore analyzing users' behaviour is an important part of web page design. Web Usage Mining (WUM) is the application of datamining techniques to web usage log repositories in order to discover the usage patterns that can be used to analyze the user's navigational behavior [1]. WUM contains three main steps: preprocessing, knowledge extraction and results analysis. The goal of the preprocessing stage in Web usage mining is to transform the raw web log data into a set of user profiles. Each such profile captures a sequence or a set of URLs representing a user session.

This sessionized data can be used as the input for a variety of data mining tasks such as clustering [2], association rule mining [3], sequence mining [4] etc. If the data mining task at hand is clustering, the session files are filtered to remove very small sessions in order to eliminate the noise from the data [5]. But direct removal of these small sized sessions may result in loss of a significant amount of information especially when the number of small sessions is large. We propose a "Fuzzy Set Theoretic" approach to deal with this problem. Instead of directly removing all the small sessions below a specified threshold, we assign weights to all the sessions using a "Fuzzy Membership Function" based on the number of URLs accessed by the sessions. After assigning the weights we apply a "Fuzzy c-Mean Clustering" algorithm to discover the clusters of user profiles. In this paper, we discuss our methodology to preprocess the web log data including data cleaning, user identification and session identification. We also describe our methodology to perform feature selection (or dimensionality reduction) and session weight assignment tasks. Finally we compare our soft computing based approach of session weight assignment with the traditional hard computing based approach of small session elimination.

## I. INTRODUCTION

Due to the digital revolution and advancements in computer hardware and software technologies, digitized information is easy to capture and fairly inexpensive to store [6], [7]. As a result huge amount of data have been collected and stored in databases. The rate at which such data is stored is growing at a phenomenal rate. The fast growing tremendous amount of data collected and stored in large and numerous data repositories, has far exceeded our human ability for comprehension without powerful tools. The abundance of data, coupled with the need for powerful data analysis tools has been described as a "data rich but information poor" situation. Hence, there is an urgent need for a new generation of computational techniques and tools to assist humans in extracting useful information (knowledge) from the rapidly growing volumes of data [8]. Data mining is the process of exploration and analysis, by automatic or semi-automatic means, of large quantities of data in order to discover meaningful patterns or rules. It deals with the "knowledge in the database" [8]. The term KDD refers to the overall process of knowledge discovery in databases. Data mining is a particular step in this process, involving the application of specific algorithms for extracting patterns from data. The additional steps in the KDD process, such as data preparation, data selection, data cleaning, incorporation of appropriate prior knowledge, and proper interpretation of the results of mining, ensures that useful knowledge is derived from the data [9].

Data mining often builds on an interdisciplinary bundle of specialized techniques from fields such as statistics, artificial intelligence, machine learning, data bases, pattern recognition, computer-based visualization etc. The more common model functions in current data mining practice include classification, regression clustering, rule generation, discovering association, summarization and sequence analysis [10]. The World Wide Web as a large and dynamic information source, that is structurally complex and ever growing, is a fertile ground for data mining principles or Web Mining. Web mining is primarily aimed at deriving actionable knowledge from the Web through the application of various data mining techniques [11]. Web data is typically unlabelled, distributed, heterogeneous, semi-structured, time varying, and high dimensional. Web data can be grouped into the following categories [12]: i) Contents of actual Web pages, ii) Intra-page structures of the web pages, iii) Inter page structures specifying linkage structures between Web pages, iv) Web usage data describing how Web pages are accessed and v) User profiles which include demographic and registration information about users. Web Usage Mining is the discovery of user access patterns from Web servers [1]. Web Usage Mining analyzes results of user interactions with a Web server, including Web logs, click streams, and database transactions at a Web site or a group of related sites. Web usage mining includes clustering (e.g. finding natural groupings of users, pages etc.), associations (e.g. which URLs tend to be requested together), and sequential analysis (the order in which URLs tend to be accessed) [13]. As with any knowledge, discovery and data mining (KDD) process, WUM performs three main steps: preprocessing, pattern extraction and results analysis. Figure 1 describes the WUM process.

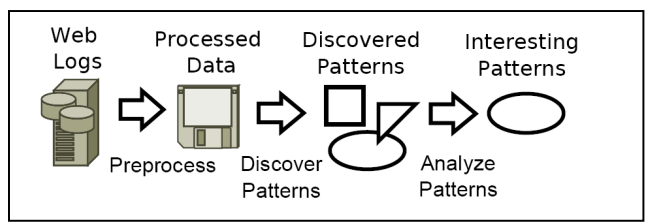

Figure 1.  Web Usage Mining Process.

The goal of the preprocessing stage in Web usage mining is to transform the raw click stream data into a set of user profiles. Each such profile captures a sequence or a set of URLs representing a user session. Web usage data preprocessing exploit a variety of algorithms and heuristic techniques for various preprocessing tasks such as data fusion and cleaning, user and session identification etc. Figure 2 depicts the primary tasks involved in web log data preprocessing in order to discover the user sessions.

Data fusion refers to the merging of log files from several Web servers. This requires global synchronization across these servers [14]. Data cleaning involves tasks such as, removing extraneous references to embedded objects, style files, graphics, or sound files, and removing references due to spider navigations. Popular Web sites generate the log file of the size measured in gigabytes per hour. Manipulating such large files is a complicated task. By filtering out useless data, we can reduce log file size to enhance the upcoming mining tasks.

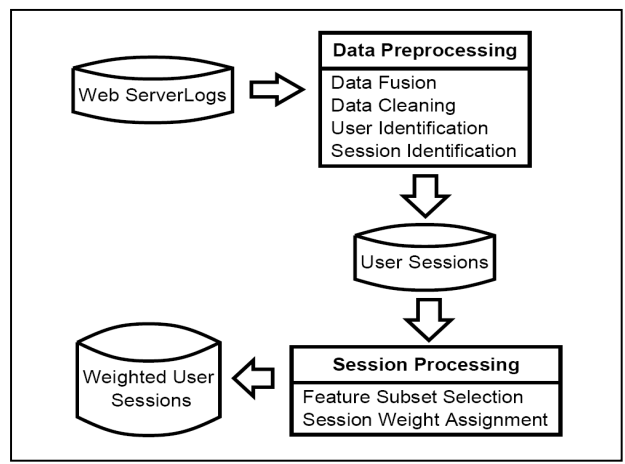

Figure 2.  Web Log Processing to Discover Weighted Sessions.

User identification refers to the process of identifying unique users from the user activity logs. Usually the log file in Extended Common Log format provides only the computer's address and the user agent. For Web sites requiring user registration, the log file also contains the user login. In such cases this information can be used for user identification. For those cases where user login information is not available, we consider each IP as a user. User Session identification is the process of segmenting the user activity log of each user into sessions, each representing a single visit to the site. Identification of user sessions from the web log file is a complicated task, due to the existence of proxy servers, dynamic addresses, and cases of multiple users access the same computer [23][2][25][26]. It is also possible that one user might be using multiple browsers or computers. This sessionized data can be used as the input for a variety of data mining algorithms.

Once user sessions are discovered, this sessionized data can be used as the input for a variety of data mining tasks such as clustering, association rule mining, sequence mining etc. If the data mining task at hand is clustering, the session files are filtered to remove very small sessions in order to eliminate the noise from the data. But direct removal of these small sized sessions may result in loss of a significant amount of information especially when the number of small sessions is large. We propose a "Fuzzy Set Theoretic" approach to deal with this problem. Instead of directly removing all the small sessions below a specified threshold, we assign weights to all the sessions using a "Fuzzy Membership Function" based on the number of URLs accessed by the sessions. After assigning the weights we apply a "Fuzzy c-Mean Clustering" algorithm to discover the clusters of user profiles. Fuzzy clustering techniques perform non-unique partitioning of the data items where each data point is assigned a membership value for each of the clusters. This allows the clusters to grow into their natural shapes [15]. A membership value of zero indicates that the data point is not a member of that cluster. A non-zero membership value shows the degree to which the data point represents a cluster. Fuzzy clustering algorithms can handle the outliers by assigning them very small membership degree for the surrounding clusters. Thus fuzzy clustering is more robust method for handling natural data with vagueness and uncertainty.

Rest of the paper is organized as follows: in section-II, we describe the techniques to preprocess the web log data including data cleaning, user and session identification. In Section III, we describe our methodology for feature selection (or dimensionality reduction) and session weight assignment. In this section we also discuss our work to apply Fuzzy c-Mean Clustering algorithms to weighted user sessions. Section IV provides the experimental results of our methodology applied to a real Web site access logs. Finally section V discusses the conclusion and future work.

## II. PREPROCESSING OF WEB LOG DATA

The primary data sources used in Web usage mining are the server log files, which include Web server access logs and application server logs.

```
1212265085.247  741  192.168.23.62  TCP_MISS/200  10858  GET
http://www.pace.edu.in/index.php - DEFAULT_PARENT/192.168.20.1
                         Mozilla/5.0
```

Figure 3. A Sample Web Log Entry.

A sample web server log file entry in Extended Common Log Format (ECLF) is given in Figure 3 and description of various fields is given in Table I.

TABLE I. DESCRIPTION OF LOG FIELDS

| Field Value | Description |
|---|---|
| 1212265085.247 | The time of request, in coordinated universal time |
| 741 | The elapsed time for HTTP request |
| 192.168.23.62 | IP address of the client |
| TCP_MISS/200 | HTTP reply status code |
| 10858 | Bytes sent by the server in response to the request. |
| GET | The requested action |
| http://www.pace.edu.in/index.php | URI of the object being requested |
| - | client user name, If disabled, it is logged as - |
| DEFAULT_PARENT/192.168.20 | Hostname of the machine where we got the object. |
| - | Content Type of the object |

### A. Data Cleaning

A user's request to view a particular page often results in several log entries since graphics and scripts are down-loaded in addition to the HTML file. In most cases, only the log entry of the HTML file request is relevant and should be kept for the user session file. This is because, in general, a user does not explicitly request all of the graphics that are on a Web page, they are automatically downloaded due to the HTML tags. Since the main purpose of Web Usage Mining is to get a picture of the user's behavior, it does not make sense to include file requests that the user did not explicitly request. During the Data cleaning process we removed the extraneous references to embedded objects, style files, graphics and sound files. Elimination of the irrelevant items was accomplished by checking the suffix of the URL name. All log entries with filename suffixes such as, gif, jpeg, GIF, JPEG, jpg, JPG, and map were removed. Default list of suffixes were used to remove undesired files. Another main activity of the cleaning process is removal of robots' requests. Web Robots or spiders scan a Web site to extract its content. Web robots automatically access all the hyperlinks from a Web page. The number of requests from a web robot is at least the number of the site's URLs. Removing WR-generated log entries removes uninteresting sessions from the log file and simplifies subsequent the mining tasks. In order to identify WR hosts we used as list of all user agents known as robots as suggested by [16]. We obtained this list from the site "http://www.robotstxt.org". Figure 4 describes the algorithm for data cleaning and transformation.

---

Input: Access log file *W*

Output: Cleaned file *C*

For each line *L ε W* do

1) Split *L* and extract various fields
2) If the *URL* includes the query string then remove it
3) Remove all the irrelevant requests whose URL suffix specified in the irrelevant suffix list
4) Remove all WR-generated requests
5) Encrypt *IP* address to hide user's identity
6) Store *URL* in a URL map along with corresponding URL number
7) Print required fields in to the output file

---

Figure 4. A Sample Web Log Entry.

Table II describes the format of the output file C generated as a result of cleaning and transformations of the web logs. The output file shows that client IP addresses are replaced with aliases in order to hide the identity of the user. The URL column of the table shows that URL strings are replaced by numbers in order to enhance further processing. We maintain a map of URL strings and corresponding URL numbers.

TABLE II. FILE FORMAT AFTER DATA CLEANING

| Time | IP | User Agent | Elapsed Time | Bytes | URL |
|---|---|---|---|---|---|
| 20080601014805 | IP1 | UA1 | 741 | 10858 | 1 |
| 20080601014806 | IP1 | UA1 | 1735 | 19247 | 2 |
| 20080601014808 | IP2 | UA2 | 239 | 209 | 1 |
| 20080601014809 | IP1 | UA3 | 674 | 156 | 3 |
| 20080601014813 | IP2 | UA2 | 680 | 179 | 4 |

## B. User Identification

Once web log files have been cleaned, next step in the data preparation is the identification of the user. Since the log files of web server we are working on do not contain the user login information, we consider each unique IP and User-Agent combination as a separate user. Next we separate out all the requests corresponding to each individual user. Figure 5 describes the algorithm to generate requests corresponding to each individual user.

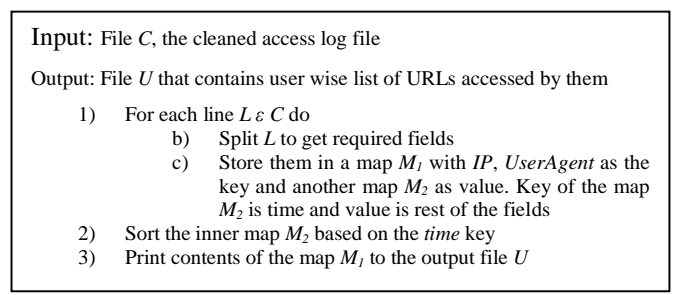

Input: File *C*, the cleaned access log file

Output: File *U* that contains user wise list of URLs accessed by them

1) For each line *L ε C* do
    b) Split *L* to get required fields
    c) Store them in a map $M_1$ with *IP*, *UserAgent* as the key and another map $M_2$ as value. Key of the map $M_2$ is time and value is rest of the fields
2) Sort the inner map $M_2$ based on the *time* key
3) Print contents of the map $M_1$ to the output file *U*

Figure 5. Algorithm to separate requests for each individual user

The format of the output file *U* generated after user identification is depicted in Table III below:

TABLE III.   FILE FORMAT AFTER USER IDENTIFICATION

| User | Time | Elapsed Time | Bytes | URL |
|---|---|---|---|---|
| U1 | 20080601014805 | 741 | 10858 | 1 |
|  | 20080601014806 | 1735 | 19247 | 2 |
|  | … | … | … | … |
| U2 | 20080601014809 | 674 | 156 | 3 |
|  | … | … | … | … |
| U3 | 20080601014808 | 239 | 209 | 1 |
|  | 20080601014813 | 680 | 179 | 4 |

## C. User Session Identification

User Session identification is the process of segmenting the user activity log of each user into sessions, each representing a single visit to the site. Web sites without user authentication information mostly rely on heuristics methods for sessionization. The sessionization heuristic helps in extracting the actual sequence of actions performed by one user during one visit to the site. In order to identify user sessions we experimented with two different time oriented heuristics (*TOH*) as described below:

- $TOH_1$ : The time duration of a session must not exceed a threshold *β*. Let timestamp of the first URL request in a session is $T_1$. A URL request with timestamp $T_i$ is assigned to this session if and only if $T_i - T_1 \leq \beta$. The first URL request with timestamp larger than $T_1 + \beta$ is considered as the first request of the next session.

- $TOH_2$: The time spent on a page visit must not exceed a threshold *β*. Let $T_i$ be the timestamp of the URL most recently assigned to a session. The next URL request with timestamp $T_{i+1}$ belongs to the same session if and only if $T_{i+1} - T_i \leq \beta$. Otherwise, this URL is considered to be the first of the next session.

Time-oriented heuristic $TOH_1$ uses an upper bound on the time spent in the entire site during a visit. The timestamp of every URL access request is compared with that of the first access request of the current session. If the time difference is larger than *β*, this request becomes the first request of the new session; otherwise it belongs to the current session. On the other hand Time-oriented heuristic $TOH_2$ uses an upper bound on page-stay time. The timestamp of every URL access request is compared with that of the previous access request. If the time difference is larger than *β*, this request becomes the first request of the new session; otherwise it belongs to the current session. We have selected 30 minutes as the value of threshold time *β* for both of the above schemes.

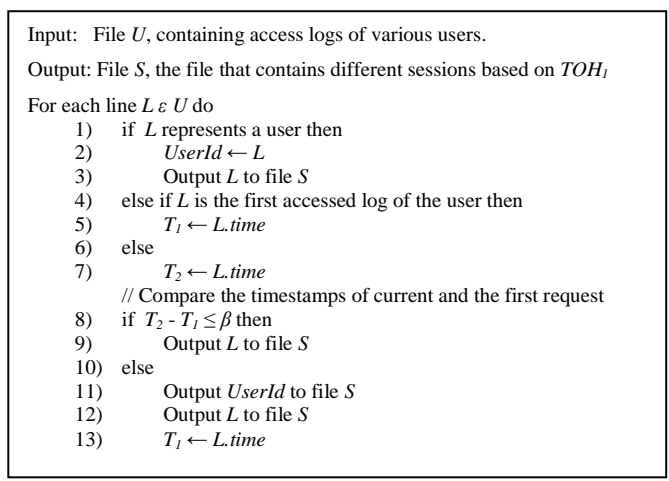

Input:   File *U*, containing access logs of various users.

Output: File *S*, the file that contains different sessions based on $TOH_1$

For each line *L ε U* do
1) if *L* represents a user then
2)     *UserId ← L*
3)     Output *L* to file *S*
4) else if *L* is the first accessed log of the user then
5)     $T_1$ ← *L.time*
6) else
7)     $T_2$ ← *L.time*
   // Compare the timestamps of current and the first request
8)     if $T_2 - T_1 \leq \beta$ then
9)         Output *L* to file *S*
10)    else
11)        Output *UserId* to file *S*
12)        Output *L* to file *S*
13)        $T_1$ ← *L.time*

Figure 6. Algorithm to generate User Sessions based on $TOH_1$

Algorithm to generate the users sessions based on the time oriented heuristics $TOH_1$ is specified in Figure 6.

TABLE IV.   FILE FORMAT AFTER USER SESSION IDENTIFICATION

| User Session | Time | Elapsed Time | Bytes | URL |
|---|---|---|---|---|
| U1-S1 | 20080601014805 | 741 | 10858 | 1 |
|  | 20080601014806 | 1735 | 19247 | 2 |
|  | … | … | … | … |
| U1-S2 | … | … | … | … |
|  | … | … | … | … |
| … |  |  |  |  |
| … |  |  |  |  |
| U2-S1 | 20080601014809 | 674 | 156 | 3 |
|  | … | … | … | … |
| U3-S1 | 20080601014808 | 239 | 209 | 1 |
|  | 20080601014813 | 680 | 179 | 4 |

Table IV shows the format the of the output file S containing user sessions. Once user sessions are generated we scan each session and remove the duplicate URLs from each session. For each unique URL within a user session a single copy of the URL is kept along with it's frequency of occurrence. We also maintain the count of the total number of unique URLs in each session.

## III. DISCOVERY OF USER SESSION CLUSTERS

### A. Feature Subset Selection of User Sessions

Each user session can be thought of a single transaction of many URL references. We map the user sessions as vectors of URL references in a n-dimensional space. Let $U$ be a set of n unique URLs appearing in the preprocessed log then $U = \{u_1, u_2, \ldots, u_n\}$ and let $S$ be a set of m user sessions discovered by preprocessing the web log data. Then $S = \{s_1, s_2, \ldots, s_m\}$ where each user session $s_i \in S$ can be represented as a bit vector $s = \{w_{u_1}, w_{u_2}, \ldots, w_{u_m}\}$ where $w_{u_i} = 1$ if $w_{u_i} \in s$, and $w_{u_i} = 0$ otherwise.

Instead of binary weights, feature weights can also be used to represent a user session. These feature weights may be based on frequency of occurrence of a URL reference within the user session, the time a user spends on a particular page or the number of bytes downloaded by the uses from a page. However, the URLs appearing in the access logs and could number in the thousands. Distance-based clustering methods often perform very poor when dealing with very high dimensional data. Therefore filtering the logs by removing references to low support URLs (i.e. that are not supported by a specified number of user sessions) can provide an effective dimensionality reduction method while improving clustering.

### B. Assiging Weights to User Sessions

If the data mining task at hand is clustering, the session files can be filtered to remove very small sessions in order to eliminate the noise from the data [5]. But direct removal of these small sized sessions may result in loss of a significant amount of information especially when the number of small sessions is large. We propose a "Fuzzy Set Theoretic" approach to deal with this problem. Instead of directly removing all the small sessions below a specified threshold, we assign weights to all the sessions using a "Fuzzy Membership Function" based on the number of URLs accessed by the sessions.

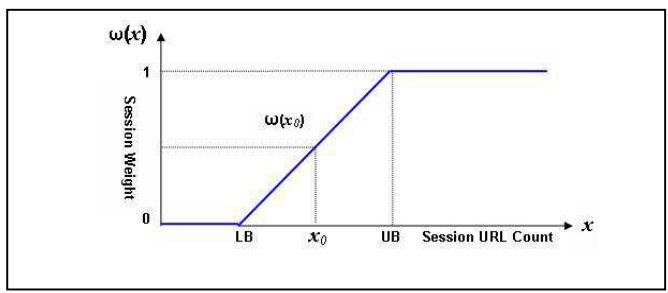

Figure 7. Fuzzy membership function for session weight assignment

Figure 7 depicts a linear Fuzzy membership function for session weight assignment. Here LB represents a lower bound on the number of URLs accessed in a session and UB represents an upper bound on the number of URLs accessed in a session. Let $|s_i|$ be the number of URLs accessed in session $s_i$ then the fuzzy membership function takes the following values:

$$\left. \begin{array}{l} W(s_i) = 0, \text{if } |s_i| \leq LB \\ W(s_i) = 1, \text{if } |s_i| \geq LB \\ W(s_i) = \dfrac{|s_i| - LB}{UB - LB}, \text{otherwise} \end{array} \right\}. \quad (1)$$

### C. Clustering the User Sessions

Once use sessions are represented in the form of a vector, clustering algorithm can be run against them. The goal of this process is to discover session clusters that represent similar URL access patterns. For example, two session vectors are similar if the Euclidean distance between them is short enough. Clustering aims to divide a data set into groups or clusters where inter-cluster similarities are minimized while the intra cluster similarities are maximized. Details of various clustering techniques can be found in survey articles [18][19][20]. The ultimate goal of clustering is to assign data points to a finite system of k clusters. Union of these clusters is equal to a full dataset with the possible exception of outliers.

The k-means clustering algorithm is one of the most commonly used methods for partitioning the data. This algorithm partitions a set of m objects into k clusters. The algorithm proceeds by computing the distances between a data point and each cluster center in order to assign the data item to one of the clusters so that intra-cluster similarity is high but inter-cluster similarity is low. Euclidian distance can be used as a measure to calculate the distance between various data points and cluster centers.

$$d(x_i, v_j) = \left\| \sum_{k=1}^{n} x_k^i - v_k^j \right\|^2 \quad (2)$$

where,

$x_i$ is the $i^{th}$ data point

$v_j$ is the $j^{th}$ cluster center

$d(x_i, v_j)$ is the distance between $x_i$ and $v_j$

$n$ is the number of dimensions of each data point

$x_k^i$ is the value of $k^{th}$ dimensions of $x_i$

$v_k^j$ is the value of $k^{th}$ dimensions of $v_j$

The k-means clustering first initializes the cluster centers randomly. Then each data point $x_i$ is assigned to some cluster $v_j$ which has the minimum distance with this data point. Once all the data points have been assigned to clusters, cluster centers are updated by taking the weighted average of all data points in that cluster. This recalculation of cluster centers results in better cluster center set. The process is continued until there is no change in cluster centers. Although k-means clustering algorithm is efficient in handling the crisp data which have clear cut boundaries, but in real world data clusters have ill defined boundaries and often overlapping clusters. This happens because many times the natural data suffer from Ambiguity, Uncertainty and Vagueness [21].

Fuzzy c-means clustering incorporates fuzzy set theoretic concept of partial membership and may result in the formation of overlapping clusters. The algorithm calculates the cluster centers and assigns a membership value to each data item corresponding to every cluster within a range of 0 to 1. The algorithm utilizes a fuzziness index parameter q where $q \in [1, \infty]$ [22] which determines the degree of fuzziness in the clusters. As the value of q reaches to 1, the algorithm works like a crisp partitioning algorithm. Increase in the value of q results in more overlapping of the clusters.

Let $X = \{x_i \mid i = 1 \cdots m\}$ be a set of *n*-dimensional data point vectors where *m* is the number of data points and each $x_i = \{x_1^i, x_2^i, \cdots, x_n^i\} \forall i = 1 \cdots m$. Let $V = \{x_j \mid j = 1 \cdots c\}$ represent a set of *n*-dimensional vectors corresponding to the cluster center corresponding to each of the *c* clusters and each $v_j = \{v_1^j, v_2^j, \cdots, v_n^j\} \forall j = 1 \cdots c$. Let $u_{ij}$ represent the grade of membership of data point $x_i$ in cluster *j*. $u_{ij} \in [1,0] \forall i = 1 \cdots m$ and $\forall j = 1 \cdots c$. The $n \times c$ matrix $U = [u_{ij}]$ is a fuzzy c-partition matrix, which describes the allocation of the data points to various clusters and satisfies the following conditions:

$$\left. \begin{array}{l} \sum_{j=1}^{c} u_{ij} = 1, \forall i = 1 \cdots m \\ 0 < \sum_{j=1}^{c} u_{ij} < m, \forall j = 1 \cdots c \end{array} \right\} \quad (3)$$

The performance index *J(U,V,X)* of fuzzy c-mean clustering can be specified as the weighted sum of distances between the data points and the corresponding centers of the clusters. In general it takes on the form:

$$J(U,V,X) = \sum_{j=1}^{c} \sum_{i=1}^{m} u_{ij}^q d_{ij}^2(\overline{x}_i, \overline{v}_j) \quad (4)$$

where,
$q \in [1, \infty]$ is the fuzziness index of the clustering
$d_{ij}^2(\overline{x}_i, \overline{v}_j)$ is the disatnce between $\overline{x}_i$ and $\overline{v}_j$

$$d_{ij}^2(\overline{x}_i, \overline{v}_j) = \sum_{k=1}^{n} w(x_i) \left\| \overline{x}_k^i - \overline{v}_k^j \right\|$$

$w(x_i)$ is the weight of the data point $x_i$

Minimization of the performance Index *J(U,V,X)* is usually achieved by updating the grade of memberships of data points and centers of the clusters in an alternating fashion until convergence. This performance Index is based on the sum of the squares criterion. During each of the iterations, the cluster centers are updated as follows:

$$\overline{v}_j = \frac{\sum_{i=1}^{m} u_{ij}^q \overline{x}_i}{\sum_{i=1}^{m} u_{ij}^q} \quad (5)$$

Membership values are calculated by the following formula:

$$u_{ij} = \frac{\left( \frac{1}{d_{ij}^2(\overline{x}_i, \overline{v}_j)} \right)^{1/(q-1)}}{\sum_{k=1}^{n} \left( \frac{1}{d_{ij}^2(\overline{x}_i, \overline{v}_j)} \right)^{1/(q-1)}} \quad (6)$$

In order to decide the number of optimum clusters for the data set *X* we use a validity function *S* which is the ratio of compactness to separation [22] as given below:

$$S = \frac{\sum_{j=1}^{c} \sum_{i=1}^{m} u_{ij}^2 \left\| \overline{x}_i - \overline{v}_j \right\|^2}{m \cdot \min_{l \neq k} \left\| \overline{v}_l - \overline{v}_k \right\|^2} \quad (7)$$

for each $c = c_{\min}, \cdots, c_{\max}$

Let $\Omega_c$ denote the optimal candidate at each *c* then, the solution to the following minimization problem yields the most valid fuzzy clustering of the data set.

$$\min_{c_{\min} \leq c \leq c_{\max}} \left( \min_{\Omega_c} S \right) \quad (8)$$

Clusters formed by the applications clustering algorithms represent a group of user sessions that are similar based on co-occurrence patterns of URL references. Clustering of user sessions results in a set $C = \{c_1, c_2, \ldots, c_k\}$ of clusters, where each $c_i$ is a subset of *S*, i.e., a set of user sessions. Each cluster represents a group of users with similar navigational patterns.

IV. EXPERIMRNTAL RESULTS

In order to discover the clusters that exist in user accesses sessions of a web site, we carried out a number of experiments. The Web access logs were taken from the P.A. College of Engineering, Mangalore web site, at URL http://www.pace.edu.in. The site hosts a variety of information, including departments, faculty members, research areas, and course information. The Web access logs covered a period of one month, from February 1, 2011 to March 1, 2011. There were 74,924 logged requests in total.

After performing the cleaning step the output file contains 30720 entries. Number of the site URLs with access count greater than or equal to 5 are 159. Total numbers of unique users identified are 24. Table V depicts the results of cleaning and user identification steps.

TABLE V. RESULTS OF CLEANING AND USER IDENTIFICATION

| Items | Count |
|---|---|
| Initial No of Log Entries | 74924 |
| Log Entries after Cleaning | 30720 |
| No. of site ULRs | 159 |
| No of Users Identified | 24 |

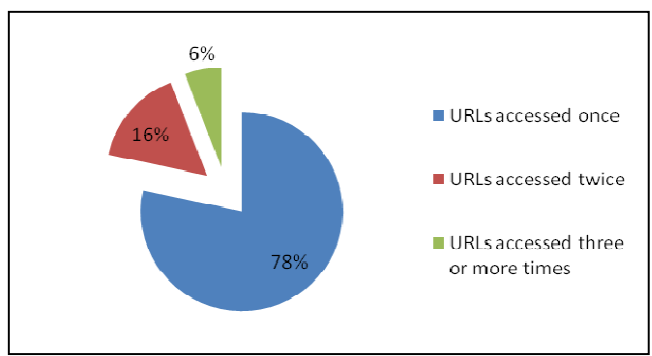

Figure 8. Percentage of URLs versus URL Access Frequency

TABLE VI. RESULTS OF CLEANING AND USER IDENTIFICATION

| Items | Count |
|---|---|
| No. of User Sessions 968 | 968 |
| Minimum no. of URLs accessed in a session | 1 |
| Maximum no. of URLs accessed in a session | 545 |
| Average no. of URLs accessed in a session | 26.12 |
| Minimum no. of unique URLs accessed in a session | 1 |
| Maximum unique URLs Accessed in a session | 158 |
| Average unique URLs Accessed in a session | 6.5 |

Total number of unique URLs of the Web Site present in the log file entries is 6850. Figure 6 shows the percentage of the URLs against how many times they are accessed in the log file. It is clear from the graph that 78% of URLs were accessed only once, 16% of them were accessed twice and only 6% of them are accessed three or more times. Maximum access count for a URL is 2234. On average each URL is accessed 4.47 times.

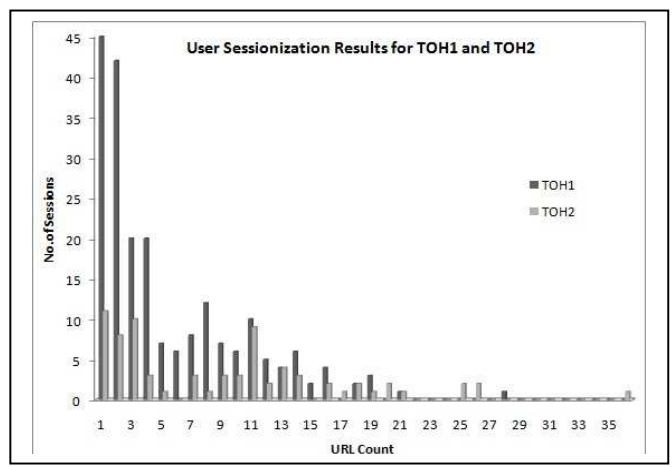

Figure 9. Sessionization results for $TOH_1$ and $TOH_2$

As far as clustering of the User Sessions is concerned those URLs which are accessed only once do not play any significant role in forming the clusters since they appear in only one of the user sessions. Therefore we eliminate all such URL requests from our further analysis. This type of URL filtering is important in removing noise from the data. Since a user session is represented by an *n*-dimensional vector, where *n* represents the number of the site URLs accessed in the log files. Reduction in the number of URLs also reduces the session vector dimensions. The count of the URLs which are accessed only once is 5372. After eliminating them the total number of unique URLs for sub sequent analysis is 1478. In order to identify the user sessions we applied two different kinds of time oriented heuristics $TOH_1$ and $TOH_2$. Details of these results and the comparisons of these approaches can be found from our previous work [17]. The result of application of $TOH_1$ is given in Table VI. Graph in Figure 9 depicts the results of application of Time oriented heuristics $TOH_1$ and $TOH_2$.

Figure 10 shows the number of URLs and their corresponding session support count. Our result shows that 396 URLs have a session support count of one. We eliminate these URLs since they can't play any significant role clusters formation. This type of session support filtering provides a form of dimensionality reduction in subsequent clustering tasks where URLs appearing in the session file are used as features. Table 4 shows the results of user session identification after the elimination of these low support URLs.

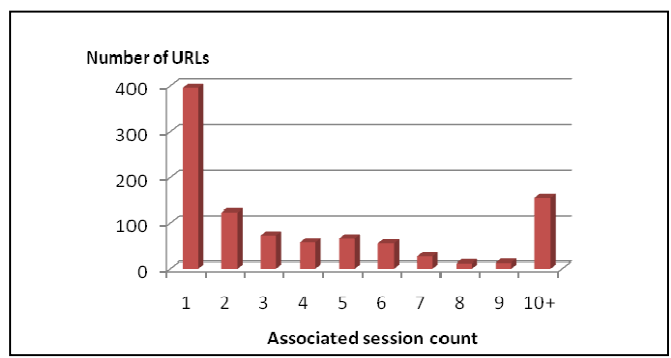

Figure 10. No. of URLs Versus No. of Sessions They are Associated with

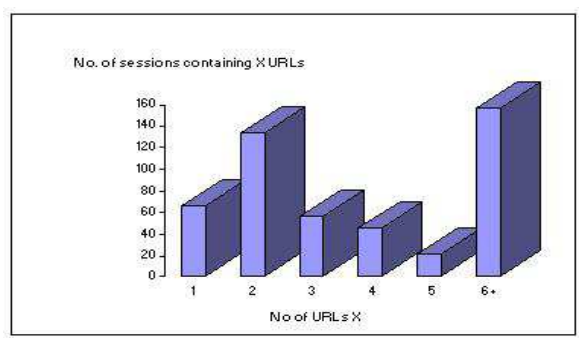

Figure 11. No. of Sessions Versus No. of URLs

Figure 11 depicts the session counts against various URL counts. Our results show that there are quite a large number of user sessions containing only few URLs. For example there are 67 sessions containing one only URL, 134 containing two URLs and 56 sessions containing three URLs. User sessions with smaller number of URLs are less significant for the purpose of clustering.

We are interested in only those sessions that access more than a certain number of URLs, say MinURLs. For example, it is not very useful to cluster user sessions which just access the URL for home page and leave. Therefore we impose certain constraints desirable for better clustering performance and outcome by using a Fuzzy set theoretic approach to assign the weights to various user sessions based on the number of URLs they contain. Instead of directly removing all the small sessions below a specified threshold, we assign weights to all the sessions using a "Fuzzy Membership Function" based on the number of URLs accessed by the sessions.

Based on the sessionization result as shown in graph of figure 11, we choose the lower bound on the number of URLs accessed in a session (LB) as 1 and an upper bound on the number of URLs accessed in a session (UB) as 6. Using equation (1) weights assigned to various sessions are specified in Table VII.

TABLE VII.  SESSION WEIGHTS BASED ON THE URL COUNT

| Session URL Count | Session Weight |
|---|---|
| 1 | 0 |
| 2 | 0.2 |
| 3 | 0.4 |
| 4 | 0.6 |
| 5 | 0.8 |
| 6 or more | 1 |

Once use sessions are assigned the weights based on the URL count, Fuzzy c-Mean clustering algorithm is applied to discover session clusters that represent similar URL access patterns. Application of the Fuzzy c-means clustering algorithm resulted in the formation of overlapping clusters. The performance Index $J(U,V,X)$ of fuzzy c-mean clustering is calculated using equation (4). It is the weighted sum of distances between the data points and the corresponding centers of the clusters. Minimization of the performance Index $J(U,V,X)$ is achieved by updating the grade of memberships of data points and centers of the clusters in an alternating fashion using the equations (6) and (5) respectively, until convergence.

Fuzzy c-Mean clustering is first applied by choosing the number of clusters as 4. During each of the iterations we increased the number of clusters by 1 till the number of clusters is reached to 60. We repeated the above process for weighted as well as non-weighted sessions. Graph is figure 12 shows the performance index (J) versus number of clusters for weighted as well as non-weighted sessions. From the graph it is clear that "Fuzzy Set Theoretic" weighted session approach results in better minimization of the performance index than non-weighted session approach.

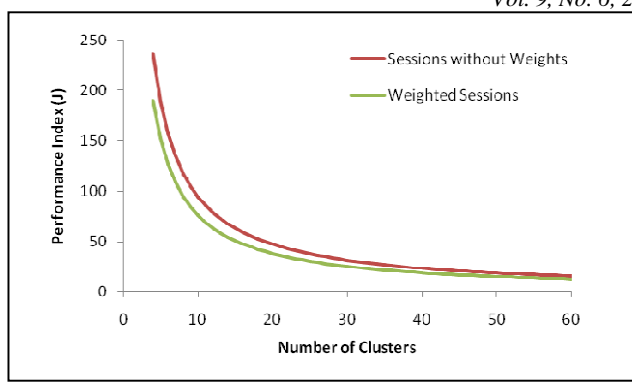

Figure 12.  No. of Clusters Versus Performance Index

In order to decide the number of optimum clusters we calculated the validity index (*S*), which is the ratio of compactness to separation using the equation (7).

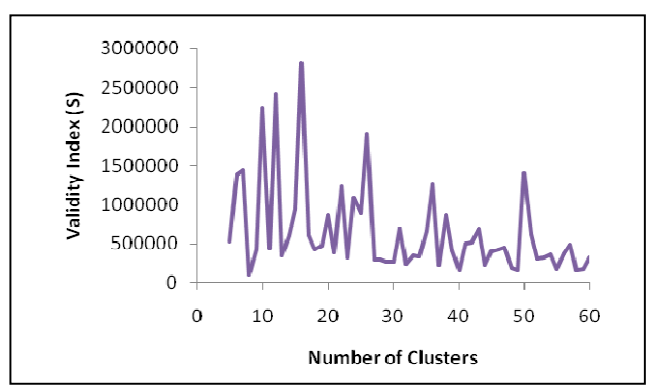

Figure 13.  Validity Index Versus No. of Clusters for Weighted Sessions

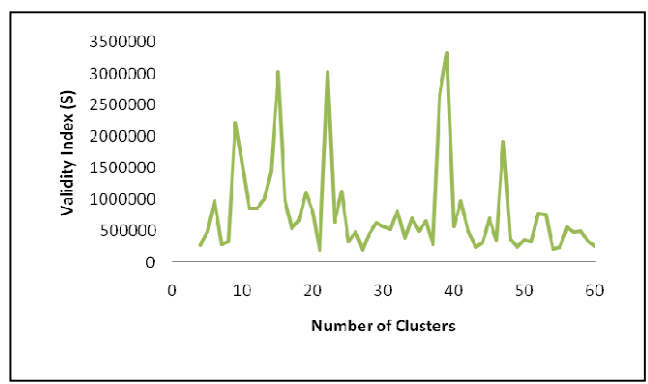

Figure 14.  Validity Index Vs. No. of Clusters for Non-Weighted Sessions

Figures 13 and 14 provide the graphs of validity index (S) versus number of clusters for weighted and non-weighted sessions respectively. Our results show that for the weighted sessions validity index is minimized when value chosen for the number of clusters is 8. On the other hand for the case of non-weighted sessions, validity index is minimized when the number of clusters is 21. Thus the optimal number of clusters for weighted sessions is 8 and for non-weighted sessions it is 21.

## V. CONCLUSION AND FUTURE WORK

In this paper, we discussed our methodology to preprocess the web log data including data cleaning, user identification and session identification. We also discussed the details about how to apply the Fuzzy c- Mean Clustering algorithm in order to cluster the user sessions.

In order improve the clustering results; we proposed a "Fuzzy Set Theoretic" approach for the removing the sessions with very few URLs. Instead of directly removing all the small sessions below a specified threshold, we assign weights to all the sessions using a "Fuzzy Membership Function" based on the number of URLs accessed by the sessions. We described our methodology to perform feature subset selection of session vectors and session weight assignment. Finally we compared our soft computing based approach of session weight assignment with the traditional hard computing based approach of small session elimination. Our results show that the "Fuzzy Set Theoretic" approach of session weight assignment results in better minimization of clustering performance index than without session weight assignment.

We believe that the above results can be further improved if we use fuzzy set theoretic approach for the inclusion of a URL in user session instead of using crisp time threshold β. In our current strategy a URL is not included in the current sessions if it comes even one second later then the specified time threshold. We can apply a similar Fuzzy set theoretic approach to the assign the weights to the URLs based on how many times they are accessed.

AUTHORS PROFILE

Zahid Ansari is a Ph.D. candidate in the Department of CSE, Jawaharlal Nehru Technical University, India. He received his ME from Birla Institute of Technology, Pilani, India. He has worked at Tata Consultancy Services (TCS) where he was involved in the development of cutting edge tools in the field of model driven software development. His areas of research include data mining, soft computing and model driven software development. He is currently with the P.A. College of Engineering, Mangalore as a Faculty. He is also a member of ACM.

Mohammad Fazle Azeem is working as Professor and Director of department of Electronics and Communication Engineering, P.A. College of Engineering, Mangalore. He received his B.E. in electrical engineering from M.M.M. Engineering College, Gorakhpur, India,

M.S. from Aligarh Muslim University, Aligarh, India and Ph.D. from Indian Institute of Technology (IIT) Delhi, India. His interests include robotics, soft computing, evolutive computation, clustering techniques, application of neuro-fuzzy approaches for the modeling, and control of dynamic system such as biological and chemical processes.

A.Vinaya Babu is working as Director of Admissions and Professor of CSE at J.N.T. University Hyderabad, India. He received his M.Tech. and PhD in Computer Science Engineering from JNT University, Hyderabad. He is a life member of CSI, ISTE and member of FIE, IEEE, and IETE. He has published more than 35 research papers in International/National journals and Conferences. His current research interests are algorithms, information retrieval and data mining, distributed and parallel computing, Network security, image processing etc.

Waseem Ahmed is a Professor in CSE at P.A. College of Engineering, Mangalore. He obtained his BE from RVCE, Bangalore, MS from the University of Houston, USA and PhD from the Curtin University of Technology, Western Australia. His current research interests include multicore/multiprocessor development for HPC and embedded systems, and data mining. He has been exposed to academic/work environments in the USA, UAE, Malaysia, Australia and India where he has worked for more than a decade. He is a member of the IEEE.